\documentclass[showpacs,amsmath,amssymb,nofootinbib,prd]{revtex4}
\usepackage{graphicx}
\usepackage{enumerate}
\usepackage{amsmath}
\usepackage{amssymb}
\usepackage{amsfonts}
\usepackage{color}

\usepackage{amsmath}
\usepackage{graphicx}
\usepackage{enumerate}

\newcommand{\beq}{\begin{equation}}
\newcommand{\eeq}{\end{equation}}
\newcommand{\bea}{\begin{eqnarray}}
\newcommand{\eea}{\end{eqnarray}}

\begin{document}
\title{RARE TOP DECAY $t \rightarrow c \gamma$ IN GENERAL THDM-III}
\author{R. Gait\'an-Lozano}
\email{rgaitan@unam.mx}
\affiliation{Departamento de F\'isica, FES-Cuautitlan,
UNAM, C.P. 54770, Estado de M\'exico, M\'exico}
\author{R. Martinez}
\email{remartinezm@unal.edu.co}
\affiliation{Departamento de F\'isica, Universidad Nacional de Colombia, Bogot\'a D.C.}
\author{J.H. Montes de Oca}
\email{jmontes@fis.cinvestav.mx}
\affiliation{Departamento de F\'isica, Centro de Investigaci\'on y de Estudios Avanzados del IPN, D.F., Mexico}

\begin{abstract}
We study the decay $t\rightarrow c\gamma$ with flavor-changing neutral interactions in scalar sector of the type III Two Higgs Doublet Model (THDM-III) with mixing between neutral scalar fields as a result of considering the most general scalar potential. The branching ratio of the decay $Br(t\rightarrow c\gamma)$ is calculated as function of the mixing parameters and masses of the neutral scalar fields.  We obtain a $Br(t\rightarrow c\gamma)$ of the order of $10^{-8}$ for the considered regions of the mixing parameters. Finally, one upper bound for the possible events is estimated to be $n=18$ by assuming a expected luminosity of the order of 300 $fb^{-1}$.
\end{abstract}
%
%
%
\maketitle
%

%
\section{Introduction}
\label{sec1}
A sensitive test for new physics are the processes of the top quark due to large mass. The predictions of the Standard Model (SM) for the top quark in flavor changing neutral (FCN) transitions are strongly suppressed \cite{eilam} as a result of the Glashow-Iliopoulos-Maiani (GIM) mechanism \cite{glashow}. However, rare decays with branching ratios (BR) of order $10^{-5}$-$10^{-6}$ may be detectable, depending on the signal. Any hint for new top quark physics at LHC would motivate further study at the next generation of collider experiments \cite{hioki}.
Recent discovery of a SM-like Higgs boson with a mass near 125-126 GeV \cite{atlas125gev,cms125gev} has generated new motivations to study the extended Higgs sector. The two Higgs doublet model (2THDM) is one of the simplest extensions of the SM, adding a second Higgs doublet with the same quantum numbers as the first one.
The versions that involve natural flavor conservation and CP conservation in the potential through the introduction of a discrete symmetry, are known as 2HDM-I \cite{2hdmI1,2hdmI2} and 2HDM-II \cite{2hdmII}. A general version which is named as 2HDM-III allows the presence of flavor-changing neutral scalar interactions (FCNSI) at a three level \cite{2hdmIII,flavor1303}. There are also some variants (known as top, lepton, neutrino), where one Higgs doublet couples predominantly to one type of fermion
\cite{BrancoReport}, while in other models it is even possible to identify a candidate for dark matter \cite{2hdm:darkmatter1,2hdm:darkmatter2}. The definition of all these models, depends on the Yukawa structure and symmetries of the Higgs sector, whose origin is still not known. The possible appearance of new sources of CP violation is another characteristic of these models \cite{2dhmCPV}.

Within 2HDM-I where only one Higgs doublet generates all gauge and fermion masses, while the second doublet only knows about this through mixing, and thus the Higgs phenomenology will share some similarities with the SM, although the SM Higgs couplings will now be shared among the neutral scalar spectrum. The presence of a charged Higgs boson is clearly the signal beyond the SM. Within 2HDM-II one also has natural flavor conservation \cite{Glashow:1976nt}, and its phenomenology will be similar to the 2HDM-I, although in this case the SM couplings are shared not only because of mixing, but also because of the Yukawa structure. The distinctive characteristic of 2HDM-III is the presence of FCNSI, which require a certain mechanism in order to suppress them, for instance one can imposes a certain texture for the Yukawa couplings \cite{Fritzsch:1977za}, which will then predict a pattern of FCNSI Higgs couplings \cite{2hdmIII}. Within all those models (2HDM I,II,III) \cite{Carcamo:2006dp}, the Higgs doublets couple, in principle, with all fermion families, with a strength proportional to the fermion masses, modulo other parameters.

In the present work, we calculate the the BR for the decay $t \rightarrow c \gamma $ in the framework of the general 2HDM.

%
\section{The general Two-Higgs-Dublet Model type III}

Given $\Phi_1$ and $\Phi_2$ two complex $SU(2)_L$ doublet scalar fields with hypercharge-one, the most general gauge invariant and renormalizable Higgs scalar
potential is
\cite{haber}
\begin{eqnarray}
V &=&m_{11}^{2}\Phi _{1}^{+}\Phi _{1}+m_{22}^{2}\Phi _{2}^{+}\Phi _{2}-\left[
m_{12}^{2}\Phi _{1}^{+}\Phi _{2}+h.c.\right]   \nonumber \\
&&+\frac{1}{2}\lambda _{1}\left( \Phi _{1}^{+}\Phi _{1}\right) ^{2}+\frac{1}{%
2}\lambda _{2}\left( \Phi _{2}^{+}\Phi _{2}\right) ^{2}  \nonumber \\
&&+\lambda _{3}\left( \Phi _{1}^{+}\Phi _{1}\right) \left( \Phi _{2}^{+}\Phi
_{2}\right) +\lambda _{4}\left( \Phi _{1}^{+}\Phi _{2}\right) \left( \Phi
_{2}^{+}\Phi _{1}\right)   \nonumber \\
&&+\left[ \frac{1}{2}\lambda _{5}\left( \Phi _{1}^{+}\Phi _{2}\right)
^{2}+\lambda _{6}\left( \Phi _{1}^{+}\Phi _{1}\right) \left( \Phi
_{1}^{+}\Phi _{2}\right) \right.   \nonumber \\
&&\left. +\lambda _{7}\left( \Phi _{2}^{+}\Phi _{2}\right) \left( \Phi
_{1}^{+}\Phi _{2}\right) +h.c.\right] ,
\end{eqnarray}
where $m_{11}^2$, $m_{22}^2$ and $\lambda_1$, $\lambda_2$, $\lambda_3$, $\lambda_4$ are real parameters and $m_{12}^2$, $\lambda_5$, $\lambda_6$, and $\lambda_7$ are complex parameters.

Now, the most general $U(1)_{EM}$-conserving vacuum expectation values are
\begin{equation}
\langle \Phi_1 \rangle= \frac{1}{\sqrt{2}}\left(
\begin{array}{c}
0 \\
v_1 \\
\end{array}
\right),
\label{vev1}
\end{equation}
\begin{equation}
\langle \Phi_2 \rangle= \frac{1}{\sqrt{2}}\left(
\begin{array}{c}
0 \\
v_2 e^{i\xi}\\
\end{array}
\right),
\label{vev2}
\end{equation}
where $v_1$ and $v_2$ are real and non-negative, $0 \leq |\xi| \leq \pi$, and
\begin{equation}
v^2 \equiv v_1^2 + v_2^2 = \frac{4 M_W^2}{g^2} = \left (246 GeV \right)^2.
\end{equation}

In Eq. (\ref{vev1}), the phase of $v_1$ is eliminated by using a global $U(1)_Y$ hypercharge rotation. In Equation (\ref{vev2}), the complex phase can be removed by redefining the complex parameters $\mu_{12}$, $\lambda_5$, $\lambda_6$, $\lambda_7$. Thus, the CP violation is explicit in the scalar potential. The neutral components of the scalar Higgs fields in the interaction basis can be written as $\Phi_a=\frac{1}{\sqrt{2}}\left( v_a + \eta_a +i \chi_a\right)$, where $\eta_a$ denote the real part. The third neutral scalar field in the interaction basis defined as $\eta_3=-\chi_1 $ $\sin\beta+\chi_2\cos\beta$ is orthogonal to the Goldstone boson for the $Z$ boson. As a result of the explicit breaking for the CP symmetry a $3\times3$ mixing matrix $R$ for fields $\eta_{1,2,3}$ is generated. This matrix relates the mass eigenstates $h_i$ with fields $\eta_i$ as follows
\begin{equation}
h_{i}=\sum_{j=1}^{3}R_{ij}\eta _{j},
\label{h-Rn}
\end{equation}
where $R$ can be written down as:
\begin{equation}
R=\left(
\begin{array}{ccc}
c_{1}c_{2} & s_{1}c_{2} & s_{2} \\
-\left( c_{1}s_{2}s_{3}+s_{1}c_{3}\right)
& c_{1}c_{3}-s_{1}s_{2}s_{3} & c_{2}s_{3} \\
-c_{1}s_{2}c_{3}+s_{1}c_{3} & -\left(
c_{1}s_{1}+s_{1}s_{2}c_{3}\right)  &
c_{2}c _{3}
\end{array}
\right)
\end{equation}
and $c_i=\cos\alpha_i$, $s_i=\sin\alpha_i$ for $-\frac{\pi}{2}\leq\alpha_{1,2}\leq\frac{\pi}{2}$ and $0\leq\alpha_3\leq\frac{\pi}{2}$. The neutral Higgs bosons $h_i$ are defined to satisfy the masses hierarchy given by the inequalities $m_{h_1}\leq m_{h_2}\leq m_{h_3}$ \cite{rot,moretti}. For the THDM with no CP violation in scalar sector the $\eta_{1}$ and $\eta_{2}$ are mixed in a $2\times2$ matrix and the mass eigenstates are CP-even while $\eta_3$ is not mixed and has CP-odd symmetry. In this case the $\eta_1$, $\eta_2$ and $\eta_3$ are equivalent to neutral scalar $H$, $h$ and psedoscalar $A$ in the 2HDM type I,II or III, respectively. By breaking the CP symmetry in Higgs sector the fields $h_{1,2,3}$ do not have well defined the CP states.
For the Yukawa interactions between fermions and scalars fields the most general structure are
\begin{eqnarray}
-\mathcal{L}_{Yukawa} &=&\sum_{i,j=1}^{3}\sum_{a=1}^{2}\left( \overline{q}%
_{Li}^{0}Y_{aij}^{0u}\widetilde{\Phi }_{a}u_{Rj}^{0}+\overline{q}%
_{Li}^{0}Y_{aij}^{0d}\Phi _{a}d_{Rj}^{0}\right.   \nonumber \\
&&\left. +\overline{l}_{Li}^{0}Y_{aij}^{0l}\Phi _{a}e_{Rj}^{0}+h.c.\right) ,
\label{yukawa}
\end{eqnarray}
where $Y_{a}^{u,d,l}$ are the $3\times 3$ Yukawa matrices. The $q_{L}$ and $l_{L}$
denote the left handed fermions doublets under $SU(2)_L$ meanwhile $u_{R}$, $d_{R}$, $l_{R}$ correspond to the right handed singlets. The zero superscript in fermions fields stands for non mass eigenstates. After getting a correct spontaneous symmetry breaking by using (\ref{vev1}) and (\ref{vev2}), the mass matrices become
\begin{equation}
M^{u,d,l}=\sum_{a=1}^{2}\frac{v_{a}}{\sqrt{2}}Y_{a}^{u,d,l},  \label{mass}
\end{equation}
where $Y_a^{f}=V_L^f Y_a^{0f}\left(V_R^{f}\right)^\dag$ for $f=u,d,l$. The $V_{L,R}^f$ matrices are used to diagonalize the fermions mass matrices and relate the physical and weak states

In order to study the rare top decay we are interested in up-type quarks fields. By using equations (\ref{h-Rn}), the interactions between neutral Higgs bosons and fermions can be written in the form of the 2HDM type II with additional contributions which arise from Yukawa couplings $Y_1$ and contain flavor change. From now on, we will omit the subscript 1 in Yukawa couplings to simplify the notation. Therefore, the interactions for up-type quarks and neutral Higgs bosons are explicitly written as
\begin{eqnarray}
\mathcal{L}_{h_{k}}^{up-quarks} &=&\frac{1}{v\sin \beta }\sum_{i,j,k}\overline{u}%
_{i}M_{ij}^{u}\left( A_{k}P_{L}+A_{k}^{\ast }P_{R}\right) u_{j}h_{k}
\nonumber \\
&&+\frac{1}{\sin \beta }\sum_{i,j,k}\overline{u}_{i}Y_{ij}^{u}\left(
B_{k}^{-}P_{L}+B_{k}^{+}P_{R}\right) u_{j}h_{k},
\label{yukawa_quarks}
\end{eqnarray}
where
\begin{equation}
A_{k}=R_{k2}-iR_{k3}\cos \beta
\label{ak}
\end{equation}
and
\begin{equation}
B_{k}^{\pm }=R_{k1}\sin \beta -R_{k2}\cos \beta \pm R_{k3}.
\label{bk}
\end{equation}
The fermion spinors are denoted as $(u_1,\,u_2,\,u_3)=(u,\,c,\,t )$. Note that Latin indices in (\ref{ak}) and (\ref{bk}) denote the three neutral Higgs bosons meanwhile Latin indices in spinors, Yukawa matrix or mass matrix are for flavor of the up-type quarks. The CP conserving case is obtained if only two neutral Higgs bosons are mixed with well-defined CP states, for instance for $\alpha_2=\alpha_3=0$ is the usual limit.
\section{Rare top decay $t\rightarrow c \gamma$}
\label{sec2}
We are interested in the contributions of the flavor changing neutral scalar interactions to the rare top decay $t\rightarrow c \gamma$ which come from previous Yukawa interactions. For the partial width of the decay $t\rightarrow c \gamma$, using Eq. (\ref{yukawa_quarks}), we have
\begin{eqnarray}
\Gamma \left( t\rightarrow c\gamma \right)  &=&\frac{\alpha G_{F}m_{t}^{3}}{%
384\pi ^{4}\sin ^{4}\beta }\left\vert Y_{ct}^{u}\right\vert ^{2}  \nonumber
\\
&&\sum_{k}\left[ \left( B_{k}^{+}\right) ^{2}\left\vert f_{1}\left( \widehat{%
m}_{k}\right) A_{i}^{\ast }+f_{2}\left( \widehat{m}_{k}\right)
A_{k}\right\vert ^{2}\right.   \nonumber \\
&&\left. +\left( B_{k}^{-}\right) ^{2}\left\vert f_{1}\left( \widehat{m}%
_{k}\right) A_{k}+f_{2}\left( \widehat{m}_{i}\right) A_{k}^{\ast
}\right\vert ^{2}\right]   \label{width}
\end{eqnarray}
where $G_F^{-1}=\sqrt{2}v^2$, $\alpha\approx1/128$ at electroweak scale and the functions $f_{1,2}$ are defined as
\begin{equation}
f_{1}\left( \widehat{m}_{k}\right) =\int_{0}^{1}dx\int_{0}^{1-x}dy\frac{
x\left( x+y-1\right) }{x^{2}+xy-\left( 2-\widehat{m}_{k}^{2}\right) x+1},
\end{equation}%
\begin{equation}
f_{2}\left( \widehat{m}_{k}\right) =\int_{0}^{1}dx\int_{0}^{1-x}dy\frac{
\left( x-1\right) }{x^{2}+xy-\left( 2-\widehat{m}_{k}^{2}\right) x+1},
\end{equation}
with $\widehat{m}_{i}=m_{h_i}/m_{t}$. In order to give the expression for branching ratio for the rare top decay we consider as an approximation to take the reported total width for top quark as $\Gamma_{\textrm{top}}\approx1.6$ GeV \cite{pdg2012}. Therefore, the branching ratio can be written as
\begin{equation}
\textrm{Br}\left( t\rightarrow c\gamma \right) =\frac{\Gamma \left( t\rightarrow
c\gamma \right) }{\Gamma _{\textrm{top}}}.
\label{br}
\end{equation}
The above expression contains too many free parameters of the model, such as the masses
Last expression contains several free parameters of the THDM, such as the masses of neutral Higgs bosons and the mixing parameter $\alpha_i$ and $\beta$. In the next section the parameters are treated to study the rare top decay $t\rightarrow c \gamma$.
%
%
%
\section{Mixing parameters and numerical results}
\label{sec3}

First we will discuss the free parameters involved in the process. The Yukawa couplings in the THDM-III are responsible for the FCNSI as shown the expression (\ref{yukawa_quarks}). One possible option to suppress these FCNSI is obtained by assuming an ansantz for the Yukawa couplings. We take into account the ansantz proposed by Cheng-Sher \cite{2hdmIII}. This ansatz assumes a specific structure for the Yukawa matrix given by $Y_{ij}^u=\sqrt{m_{i}m_{j}}/M_W$.

For the masses of neutral scalar $h_i$ we set the mass of the lightest Higgs boson $h_1$ equal to the value of the mass of the observed scalar reported by ATLAS and CMS, $m_{h_1}\approx 126$ GeV \cite{atlas125gev,cms125gev}. The masses of the $h_2$ and $h_3$ are fixed as 300 GeV and 600 GeV, respectively. If neutral scalar fields have greater values of masses, then their contribution to the $\textrm{Br}\left( t\rightarrow c\gamma \right)$ will be negligible. Therefore, the set of the free parameters in the partial width (\ref{width}) is reduced only to the mixing angles $\left\{\alpha_1,\,\alpha_2,\,\alpha_3,\,\beta\right\}$. In order to analyze the branching ratio for rare top decay $t\rightarrow c \gamma$ we consider allowed regions for the mixing parameters $\alpha_1$ and $\alpha_2$. The numerical results show that under above assumptions the branching ratio (\ref{br}) does not have significant contributions from $\alpha_3$ mixing parameter in the interval $0\leq\alpha_3\leq\pi/2$. Then, we just focus in the $\alpha_{1,2}$ parameters. The considered regions for $\alpha_1$ and $\alpha_2$ are studied in previous work by the authors \cite{euro2014}. These allowed regions for the $\alpha_{1,2}$ parameter space are obtained by experimental and theoretical constrains in the framework of the 2HDM type II with CP violation for fixed $\tan\beta$ and the mass of the charged Higgs bosons $m_{H^\pm}$ \cite{moretti}.
\begin{figure} 
\begin{minipage}{\columnwidth}
\centering
\includegraphics[scale=0.8]{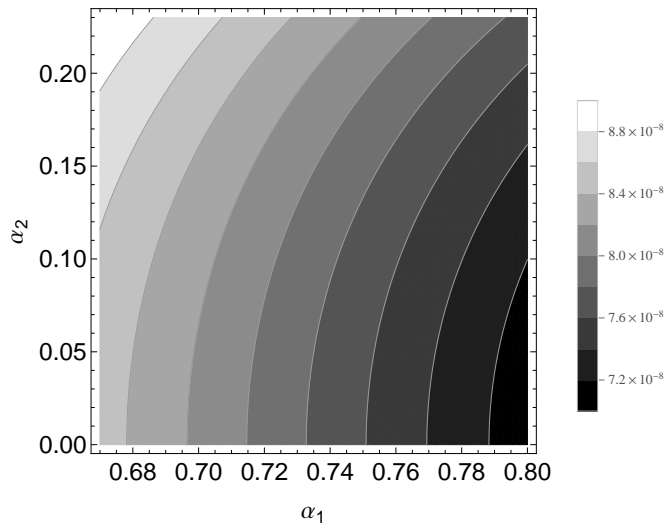}
\end{minipage}
\caption{Type III THDM branching ratio for $t\rightarrow c \gamma$ as a function of $\alpha_1$-$\alpha_2$ in regions $R_1$.}
\label{figure1}
\end{figure}
\begin{figure} 
\begin{minipage}{\columnwidth}
\centering
\includegraphics[scale=0.8]{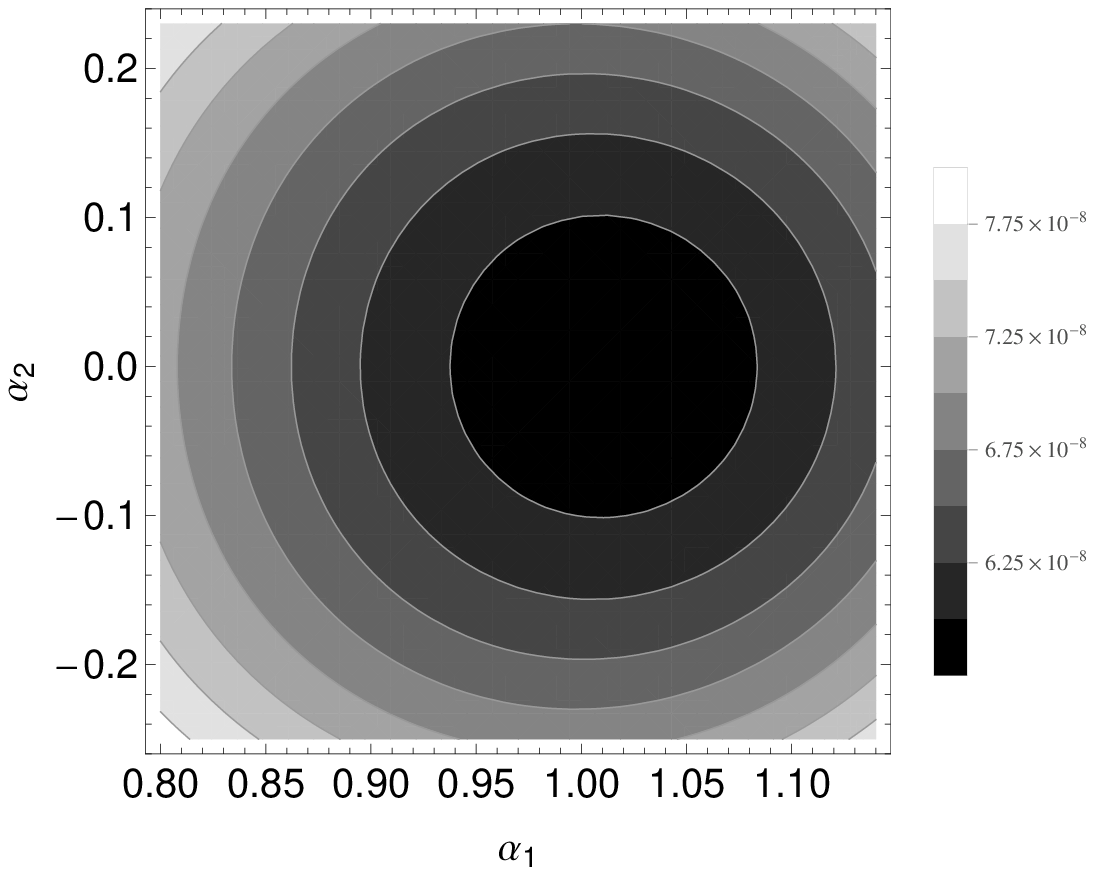}
\end{minipage}
\caption{Type III THDM branching ratio for $t\rightarrow c \gamma$ as a function of $\alpha_1$-$\alpha_2$ in regions $R_2$.}
\label{figure2}
\end{figure}
\begin{figure} 
\begin{minipage}{\columnwidth}
\centering
\includegraphics[scale=0.8]{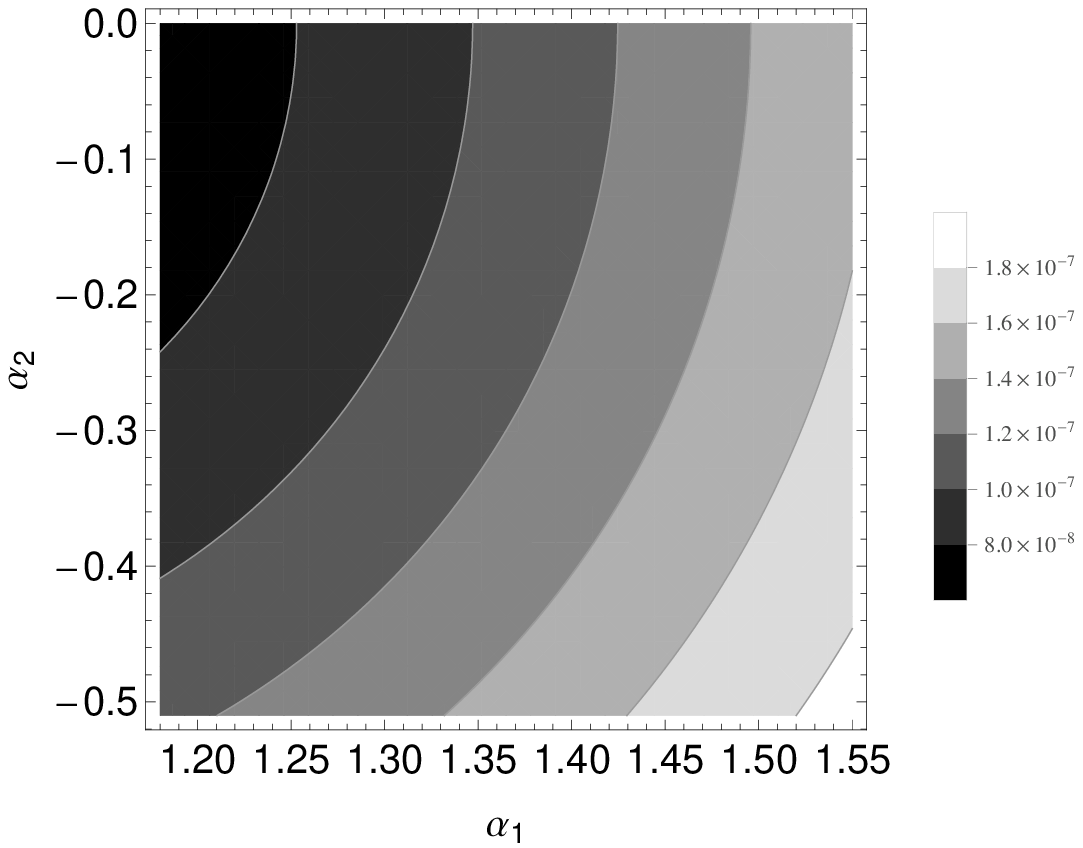}
\end{minipage}
\caption{Type III THDM branching ratio for $t\rightarrow c \gamma$ as a function of $\alpha_1$-$\alpha_2$ in regions $R_3$.}
\label{figure3}
\end{figure}
One can obtain the following regions for $\alpha_1$ and $\alpha_2$ for $0.5\leq R_{\gamma\gamma} \leq 2$, $m_{H^\pm}=300$ GeV and $\tan\beta=1$:
\begin{equation}
R_{1}=\left\{ 0.67\leq \alpha _{1}\leq 0.8\right.\,\textrm{ and}\,\left. 0\leq \alpha
_{2}\leq 0.23\right\},
\end{equation}
\begin{equation}
R_{2}=\left\{ 0.8\leq \alpha _{1}\leq 1.14\right. \,\textrm{and}\,\left. -0.25\leq \alpha
_{2}\leq 0\right\}.
\end{equation}
and
\begin{equation}
R_{3}=\left\{ 1.18\leq \alpha _{1}\leq 1.55\right. \,\textrm{and}\,\left. -0.51\leq \alpha
_{2}\leq 0\right\}.
\end{equation}
%
%
\begin{figure} 
\centering
\includegraphics[scale=0.35]{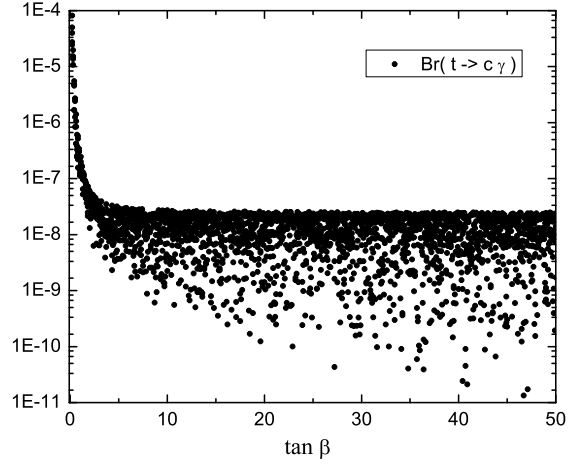}
\caption{Scatter plot for branching ratio of the rare top decay as function of $\tan\beta$ with random values for $\alpha_1$ and $\alpha_2$.}
\label{fig_sp}
\end{figure}
The figures \ref{figure1}, \ref{figure2} and \ref{figure3} show the behavior of the branching ratio as function of $\alpha_1$ and $\alpha_2$ in the allowed regions $R_1$, $R_2$ and $R_3$, respectively. These regions are restrictive for $\beta$ mixing parameter. In order to explore the behavior of the branching ratio for $\beta$ mixing parameter in an greater range we generate a set of random values for $\alpha_1$ and $\alpha_2$ and obtain the figure \ref{fig_sp}, which shows an accumulation of points in the values of $10^{-8}\sim10^{-7}$ for the branching ratio. We note that the contributions from FCNSI are greater than SM contributions \cite{mele}.

\section{Conclusions}
\label{sec4}
\begin{figure} 
\begin{minipage}{\columnwidth}
\centering
\includegraphics[scale=0.75]{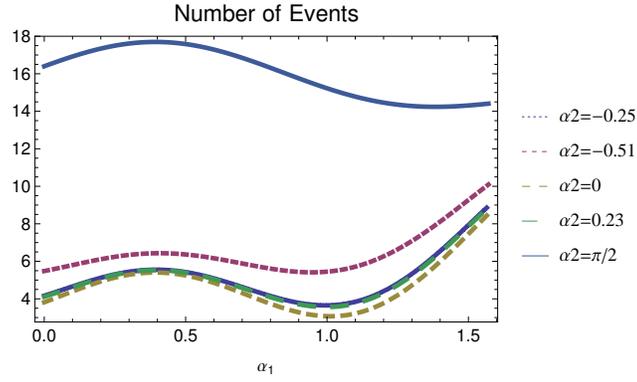}
\end{minipage}
\caption{Events for $\alpha_2$ based in the regions $R_{1,2,3}$ in the Run III of LHC.}
\label{event}
\end{figure}

From 2015 to 2017 the experiment is expected to reach 100 $\textrm{fb}^{-1}$ of data with a energy of the center of mass of 14 TeV. In the year 2021 is expected to reach a luminosity of the order of 300 $\textrm{fb}^{-1}$ of data. Experiments with this luminosity could find evidence of new physics beyond SM. Then, Run 3 in LHC could observe events for the flavor changing neutral processes, which can be explained in a naive form as $\textrm{Br}(p\bar{p}\rightarrow\bar{b}Wc\gamma)\approx \sigma (p\bar{p}\rightarrow t\bar{t})\textrm{Br}(\bar{t}\rightarrow \bar{b}W)\textrm{Br}(t\rightarrow c\gamma)$.
We estimate the number of events using the expected luminosity of $300\,\textrm{fb}^{-1}$ and $\sigma(p\bar{p}\rightarrow t\bar{t})\approx 176 \,pb $ \cite{pdg2012}. Under these assumptions, the figure \ref{event} shows the number of possible events for $t\rightarrow c \gamma$ as function of $\alpha_1$ in the allowed regions $R_{1,2,3}$.
%
%
\begin{figure} 
\begin{minipage}{\columnwidth}
\centering
\includegraphics[scale=0.7]{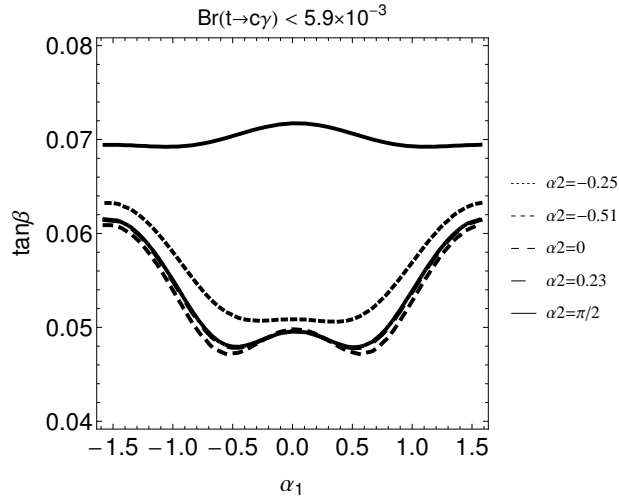}
\end{minipage}
\caption{Solution to the equation $\textrm{Br}\left( t\rightarrow c\gamma \right)=5.9\times 10^{-3}$ for $\alpha_2$ based in the regions $R_{1,2,3}$.}
\label{figure5}
\end{figure}
Last experimental results have obtained a bound for these branching ratio such as $Br(t\rightarrow c \gamma)<5.9\times10^{-3}$ \cite{pdg2012}. If we fix the branching ratio (\ref{br}) equal to the experimental upper bound, then lower bound for the $\beta$ parameter is constrained $0.048\,\leq\tan\beta$ for any $\alpha_1$ and the values of $\alpha_2$ from $R_{1,2,3}$, see Fig. \ref{figure5}. We note that the branching ratio decreases as increase the value of $\tan\beta$. In the case of the scatter plot \ref{fig_sp}, we can estimate from 1 to 5 possible events for $\textrm{Br}(t\rightarrow c \gamma)$ from $10^{-8}$ to $10^{-7}$.
%
%
%
\section*{Acknowledgments}
This work is supported in part by PAPIIT project IN117611-3, Sistema
Nacional de Investigadores (SNI) in M\'exico. J.H. Montes de Oca is thankful for support from the postdoctoral CONACYT grant and part of this work was done in his post-doctorate in UNAM with DGAPA grant. R. M. thanks to COLCIENCIAS for the financial support.
%
%
%
%

\end{document}